\documentstyle[pra,aps,multicol]{revtex}

\newcommand{\be}{\begin{equation}}
\newcommand{\ee}{\end{equation}}

\begin{document}
\draft
\title{Exact coherent states in one-dimensional quantum many-body systems 
with inverse-square interactions}
\author{Dae-Yup Song\footnote[1]{Electronic address: 
        dsong@sunchon.ac.kr}}
\address{ Department of Physics, Sunchon National University, Sunchon 
540-742, Korea}
\date{\today}
\maketitle
\begin{abstract}
For the models of $N$-body identical harmonic oscillators interacting through 
potentials of homogeneous degree -2, the unitary operator that transforms a system
of time-dependent parameters  into that of unit spring constant and unit mass of 
different timescale is found. If the interactions can be written in terms of 
the differences between positions of two particles, it is also shown that
the Schr\"{o}dinger equation is invariant under a unitary transformation. 
These unitary relations can be used not only in finding coherent states 
from the given stationary states in a system, but also in finding exact wave 
functions of the Hamiltonian systems of time-dependent parameters from those of 
time-independent Hamiltonian systems. Both operators are invariant under the 
exchange of any pair of particles.
The transformations are explicitly applied for some of the Calogero-Sutherland 
models to find exact coherent states.
\end{abstract}
\pacs{03.65.Ge, 03.65.Ca, 05.30.-d, 03.65.Fd}

\begin{multicols}{2}
\section{Introduction}
The harmonic oscillator (of time-dependent parameter) is a model where 
the path integral for the kernel (propagator) is Gaussian and thus the kernel 
can be almost determined by the classical action. From the fact that the kernel 
should satisfy Schr\"{o}dinger equation, with a condition for the kernel in the 
coincident limit, one can obtain the exact expression of the kernel in terms of 
the solutions of classical equation of motion \cite{Song}. This is one of the 
basic reasons of the fact that wave functions of harmonic oscillators are 
described by the solutions of classical equation of motion. Two different types 
of operators have been known for the construction of coherent states from the 
vacuum state of a simple harmonic oscillator system; the displacement operator 
and squeeze operator \cite{Sch,Coherent,squeeze}. 
Recently, it has been shown \cite{SongUni} that a harmonic oscillator of 
time-dependent parameters are unitarily equivalent to a simple harmonic oscillator 
of different timescale (including the case with an inverse-square potential
\cite{Cal3body}), where the unitary operator of the relation is again given 
in terms of the classical solutions of the system of time-dependent parameters. 
If one considers the set of wave functions whose centers of the probability 
distribution functions do not move, the operator at a given time corresponds to a 
squeeze operator. As have been well known, a driven harmonic oscillator 
(without an inverse-square potential) is unitarily equivalent to a harmonic 
oscillator without driving force, and this relation can be used in finding the 
unitary transformation which does not change the form of the Schr\"{o}dinger 
equation in a harmonic oscillator. The unitary operator for this transformation 
corresponds to a displacement operator. By applying this transformation to the 
eigenstates of stationary probability distribution of the model, one can obtain 
the wave functions whose centers of the probability distributions move according 
to the the classical solution.  

In this paper, we will consider the models of identical $N$-body harmonic 
oscillators of time-dependent parameters interacting through the potential 
$V(x_1,x_2,\cdots,x_N)$ of homogeneous degree -2 which satisfies
\be
V(a x_1,ax_2,\cdots,ax_N)= a^{-2}V(x_1,x_2,\cdots,x_N)
\ee
with non-zero constant $a$.
One of the models which belongs to this category was first solved by 
Sutherland \cite{Sutherland}, based on the earlier work by Calogero 
\cite{Calogero}. From its inception, this model is closely related to the random 
matrix model \cite{Mehta,SLA} and has been found relevant for the descriptions 
of various physical phenomena \cite{SLA}. The model has been generalized into the 
several cases \cite{Sutherland2,Sutherland3,Wolfes}, known in the literature as the
Calogero-Sutherland models, which have generated wide interest \cite{HS,TSA}, 
while some of the generalized models do not belong to the category we will consider
(see Sec. V.).

We will show that, there is a unitary transformation which relates the system to 
the the model of different timescale with unit mass and spring constant. 
The operator for the transformation is given as a product of 
the squeeze-type unitary operators in (one-body) harmonic oscillators. 
If $V(x_1,x_2,\cdots,x_N)$ is written in terms of the differences between positions 
of two particles, so that
\be
V(x_1+a,x_2+a,\cdots,x_N+a)=V(x_1,x_2,\cdots,x_N),
\ee
there also exists a unitary transformation which does not change the form of the 
Schr\"{o}dinger equation, while the operator for the transformation is given as the 
product of the displacement-type unitary operators in (one-body) harmonic oscillators. 
Both unitary operators are symmetric under the exchange of any pair of particles. 
These operators, as in the one-body harmonic oscillator case \cite{SongUni}, 
thus can be used to find the wave functions of the systems of time-dependent 
parameters from those of constant parameters, preserving the symmetric property of 
the wave functions. 
The unitary transformations will be explicitly applied for three cases in the 
Calogero-Sutherland models, to find exact wave functions.
In a Sutherland model of time-dependent parameters, Sutherland \cite{SCoherent}
has found an exact "coherent" state by directly analyzing the Schr\"{o}dinger equation; 
we will show that this "coherent" state is obtained by applying the squeeze-type 
operator, while the displacement-type operator can also be applied in this model 
to give a more general form of the exact wave function. In addition, by extending the 
definition of the displacement-type operator, we will also show that one can find a
unitary relation between the system without external force and the same system
with external force.  
Since the quantum states we will find through unitary transformations can be solely
described, up to some parameters of the models, by the classical solutions of a harmonic 
oscillator, we will call them coherent states.  

This paper will be organized as follows;
in the next section, we will introduce the squeeze-type unitary operator which 
relates the interacting $N$-body oscillator 
system of time-dependent parameters and the same system of constant 
parameters. The displacement-type operator will be also introduced, and it will be 
shown that, if the potential of interaction satisfies Eq. (2), the unitary 
transformation by the operator does not change the form of the Schr\"{o}dinger equation. 
In Sec. III, the unitary transformations will be explicitly applied for three cases 
in the Calogero-Sutherland models. In Sec. IV, for the cases that the potential of the 
interaction satisfies Eq. (2), it will be shown that the displacement-type operator 
can be extended to give the unitary relation between a system without external force 
and the same system with external force. The last section will be devoted 
to a summary and discussions.

\section{The unitary transformations}
It has been shown that \cite{SongUni}, the harmonic oscillator with inverse-square 
potential described by the Hamiltonian 
\be
H_{s,1}={p^2 \over 2}+{x^2 \over 2}+{g \over x^2}
\ee 
with a coupling constant $g$,
and the oscillator with time-dependent mass $M(t)$, spring constant $w(t)$ 
described by the Hamiltonian
\be
H_1={p^2\over 2M(t)}+{1\over 2}M(t)w^2(t)x^2+{g\over M(t)}{1\over x^2} 
\ee
are related through the unitary transformation. For the case of $g=0$, 
the classical equation of motion for the system of the Hamiltonian in Eq. (4) is 
written as;
\be
{d \over {dt}} (M \dot{{x}}) + M(t) w^2(t) {x} =0.
\ee
If we denote the two linearly independent solutions of Eq. (5) as $u(t)$ and $v(t)$,
the $\rho(t)$ defined by $\rho(t)=\sqrt{u^2+v^2}$ satisfies
\be
{d \over {dt}} (M \dot{\rho})-{\Omega^2 \over M\rho^3}+Mw^2 \rho=0
\ee
with a time-constant $\Omega$ ($\equiv M(t)[ \dot{v}(t)u(t) - \dot{u}(t)v(t)]$), 
while the overdots denote differentiation with respect to $t$. 
Without losing generality we assume that $\Omega$ is positive.
By defining the operator $O_{s,1}$ and $O_1$ as
\begin{eqnarray}
O_{s,1}(\tau)=-i\hbar {\partial \over \partial\tau}+ H_{s,1} \cr
O_1(t)=-i\hbar {\partial \over \partial t}+ H_{1},
\end{eqnarray}
if the $\tau$, the time of the system of $H_{s,1}$, and $t$, the time of the system 
of $H_1$, is related as 
\be
d\tau={\Omega \over \rho^2} dt,
\ee
the unitary relation between the two operators has been given in Ref. \cite{SongUni}
as
\be 
U_1 O_{s,1}(\tau)U_1^\dagger\mid_{\tau=\tau(t)}= {M \rho^2 \over \Omega} O_1,
\ee
where
\be
U_1=\exp[{i \over 2 \hbar}M {\dot{\rho} \over \rho}x^2]
\exp[-{i\over 4\hbar}\ln({\rho^2 \over \Omega})(xp+px)].
\ee
 
There is a similar unitary relation between the identical $N$-body harmonic 
oscillators interacting through the potential $V(x_1,x_2,\cdots,x_N)$ satisfying 
Eq. (1). The potential $V$ may be written as a linear combination of the terms 
${1 / \sum_{l,m=1}^N a_{lm}x_lx_m}$, where $a_{lm}=a_{ml}$.
If a system is described by the Hamiltonian
\be
H_{s,N}=\sum_{i=1}^N ({p_i^2 \over 2}+{x_i^2 \over 2})+ V(x_1,x_2,\cdots,x_N)
\ee
and another system is described by the Hamiltonian
\begin{eqnarray}
H_N &=& \sum_{i=1}^N ({p_i^2 \over 2M(t)}+M(t)w^2(t){x_i^2 \over 2})\cr
    & &   +{1 \over M(t)}V(x_1,x_2,\cdots,x_N),
\end{eqnarray}
from the commutator relation 
\begin{eqnarray}
&&[\sum_{i}^N (x_ip_i+p_ix_i), {1 / \sum_{l,m=1}^N a_{lm}x_lx_m}]\cr
&& = 4i\hbar  / \sum_{l,m=1}^N a_{lm}x_lx_m,
\end{eqnarray}
one may find the unitary relation of the two systems 
\be
U_N O_{s,N}(\tau)U_N^\dagger \mid_{\tau=\tau(t)}= {M \rho^2 \over \Omega} O_N,
\ee
where
\begin{eqnarray} 
O_{s,N}(\tau)&=&-i\hbar {\partial \over \partial\tau}
        +H_{s,N}\\
O_N(t)&=&-i\hbar {\partial \over \partial t} + H_N.
\end{eqnarray}
This relation has been noticed for a specific case \cite{BCG}.
In Eq. (14), the unitary operator $U_N$ is given as
\begin{eqnarray}
&&U_N  \cr
&&=\prod_{i=1}^N  (\exp[{i \over 2 \hbar}M {\dot{\rho} \over \rho} x_i^2]
     \exp[-{i\over 4\hbar}\ln({\rho^2 \over \Omega})(x_ip_i+p_ix_i)]) \\
&&=({\Omega \over \rho^2})^{N/4} 
     \prod_{i=1}^N (\exp[{i \over 2 \hbar}M {\dot{\rho} \over \rho} x_i^2]
     \exp[-{1 \over 2}\ln({\rho^2 \over \Omega})x_i {\partial \over \partial x_i}]).
\end{eqnarray}
If $\phi_s(x_1,x_2,\cdots,x_N)$ is an eigenstate of Hamiltonian $H_{s,N}$ with 
eigenvalue $E$, from Eqs. (14,18), the wave function 
$\psi(t;x_1,x_2,\cdots,x_N)$  satisfying $O_N(t)\psi=0$ is given as
\begin{eqnarray}
&&\psi(t;x_1,x_2,\cdots,x_N)  \cr
&&=e^{-iE\tau/\hbar}\mid_{\tau=\tau(t)}
         U_N\phi_s(x_1,x_2,\cdots,x_N)\\
&&=({\Omega \over \rho^2})^{N/4} 
     ( {u(t)-iv(t) \over \rho(t)} )^{E/\hbar}
     ( \prod_{i=1}^N\exp[{i \over 2 \hbar}M {\dot{\rho} \over \rho} x_i^2] )\cr
&&~~ \times \phi_s(\sqrt{\Omega \over \rho^2}x_1,\sqrt{\Omega \over \rho^2}x_2,\cdots,
       \sqrt{\Omega \over \rho^2}x_N).
\end{eqnarray}

If $V(x_1,x_2,\cdots,x_N)$ is written in terms of the differences of positions of 
two particles, so that $V$ satisfies Eq.(2), there is a unitary operator 
\be
U_f= e^{{i \over \hbar} N\delta_f}\prod_{i=1}^N (\exp[{i\over \hbar}M\dot{u}_fx_i]
     \exp[-{i\over \hbar}u_f p_i])
\ee
which does not change the $O_N$ under a unitary transformation:
\be
U_f O_N U_f^\dagger =O_N.
\ee
In Eq. (21), $u_f$ is a linear combination of $u(t), v(t)$, and $\delta_f$ is defined 
through the relation
\be
\dot{\delta}_f= {1 \over 2}M(w^2 u_f^2 -\dot{u}_f^2).
\ee  
Therefore, in this case, a coherent wave function from $\phi_s(x_1,x_2,\cdots,x_N)$
is given as 
\begin{eqnarray}
&&\psi^f(t;x_1,,x_2,\cdots,x_N)  \cr
&&=e^{-iE\tau/\hbar}\mid_{\tau=\tau(t)}
         U_fU_N\phi_s(x_1,x_2,\cdots,x_N)\\
&&=({\Omega \over \rho^2})^{N/4} 
     ( {u-iv \over \rho} )^{E/\hbar}e^{{i \over \hbar} N\delta_f}   \cr
&&~~ \times  ( \prod_{i=1}^N  \exp[{i \over 2 \hbar}M {\dot{\rho} \over \rho} (x_i-u_f)^2 
           +{i\over \hbar}M\dot{u}_fx_i] )\cr
&&~~ \times \phi_s(\sqrt{\Omega \over \rho^2}(x_1-u_f),\cdots,
           \sqrt{\Omega \over \rho^2}(x_N-u_f)).
\end{eqnarray}

From the derivations through unitary transformations, it is manifest that
\be 
\int \prod_{i=1}^N dx_i\phi_s^*\phi_s =\int \prod_{i=1}^N dx_i \psi^*\psi
=\int \prod_{i=1}^N dx_i {\psi^f}^*\psi^f.
\ee
Since the two unitary operators, $U_N, U_f$ are invariant under the exchange of a 
pair of $i$-th and $j$-th particles,  the wave functions $\phi_s(x_1,x_2,\cdots,x_N)$, 
$\psi(t;x_1,x_2,\cdots,x_N)$ and $\psi^f(t;x_1,x_2,\cdots,x_N)$ have the same symmetric 
property under the exchanges of particles. 
For the systems of identical particles, one of the quantities of interest which is 
independent of statistics \cite{Sutherland,SCoherent,Mehta} is the particle number 
density defined for $\phi_s$ as
\be
\sigma_s(x)= N{\int_{-\infty}^\infty dx_2 \cdots \int_{-\infty}^\infty dx_N
\phi_s^2(x,x_2,\cdots,x_N) \over
\int_{-\infty}^\infty dx_1 \cdots \int_{-\infty}^\infty dx_N
\phi_s^2(x_1,x_2,\cdots,x_N)}.
\ee 
From the Eq. (25), one can easily find the expression of the particle number density  
for $\psi^f$  as
\be
\sigma^f(x)= {\sqrt{\Omega} \over \rho} \sigma_s({\sqrt{\Omega} \over \rho} (x-u_f) ).
\ee

\section{Applications}
In this section, the general results of previous section will be explicitly 
applied for three cases. First, we will consider the Sutherland model of 
Ref. \cite{Sutherland}. We will find a general expression of a coherent state of 
the system (of time-dependent parameters) and will show that the expression 
reproduces the known state in the model \cite{SCoherent}. Second, we will 
consider the model of three-body system \cite{GKP,Wolfes}.
Third, we will consider the Calogero model "in the Jacobi coordinate" \cite{Calogero}
without the degree of freedom of center of mass.

\subsection{Sutherland model}
The system described by the Hamiltonian 
\be
H_{S,s}=\sum_{i=1}^N ({p_i^2 \over 2}+{x_i^2 \over 2})
     + \sum_{i>j=1}^N {\hbar^2 \lambda (\lambda-1) \over (x_i -x_j)^2 }
\ee
has the (unnormalized) bosonic ground state \cite{Sutherland}
\be
\phi_S= (\prod_{j>i=1}^N|x_j-x_i|^\lambda)\prod_i^N e^{-x_i^2/2\hbar}
\ee
with the energy eigenvalue $\hbar N [1+ \lambda(N-1)]/2$.
For the system described by the Hamiltonian 
\begin{eqnarray}
H_S &=& \sum_{i=1}^N ({p_i^2 \over 2M(t)}+M(t)w^2(t){x_i^2 \over 2}) \cr
    & &+{1\over M(t)} \sum_{i>j=1}^N {\hbar^2 \lambda (\lambda-1) \over (x_i -x_j)^2 },
\end{eqnarray}
the wave function $\psi_S^f$ satisfying 
\be
i\hbar{\partial\psi_S^f \over\partial t}=H_S\psi_S^f
\ee
is given, from the results of previous section, as  
\begin{eqnarray}
\psi_S^f&=&
({u+iv \over \sqrt{\Omega}})^{-N(1+\lambda(N-1))/2} 
       e^{iN\delta_f/\hbar} \cr
   &&\times(  \prod_{i=1}^N\exp[{i\over 2\hbar}M{\dot{\rho} \over \rho}(x_i-u_f)^2
          +{i\over \hbar}M\dot{u}_fx_i]  )\cr
   &&\times(\prod_{j>i=1}^N|x_j-x_i|^\lambda)
      \prod_{i=1}^N  e^{-\Omega (x_i-u_f)^2/2\hbar \rho^2}.
\end{eqnarray}
In deriving Eq. (33), we make use of the fact that $(u-iv)/\rho^2=1/(u+iv)$.
By making use of Eq. (28) and the results in Ref. \cite{SCoherent,Sutherland,Mehta}, 
one can find the particle density for $\psi_S^f$ is give as
\be
\sigma_S^f(x)= {\sqrt{2N\Omega}\over \pi\rho\sqrt{\lambda}}
     \sqrt{1-{\Omega \over 2N\lambda \rho^2}(x-u_f)^2}.
\ee

By directly analyzing the Schr\"{o}dinger equation (Eq. (32)), for the unit mass 
case, a coherent wave function is given in terms of a complex solution of classical 
equation of motion, and a hydrodynamic description was shown to hold exactly in the 
picture that the wave function provides \cite{SCoherent}. 
If we choose $u_f=0$, one can easily verify that, for the unit mass case, 
the $\psi_S^f$ and $\sigma_S^f$ reduce to the wave function and the density 
found by Sutherland \cite{SCoherent}, respectively. 

As in the case of one-body harmonic oscillator \cite{SongUni}, there are, in general, 
five free parameters in determining $\psi_S^f$ or $\sigma_S^f$; two of the parameters
determine the motion of center of the particle number density, while the other three
parameters determine the shape of the density function. To be explicit, we consider 
the particle density function of the case of unit mass and unit spring constant.
In this case, two homogeneous solutions $u(t),v(t)$ and the (fictitious) particular 
solution $u_f(t)$ can be taken, without losing generality, 
as $\cos(t+t_0), A\sin(t+\alpha+t_0)$ and $B\cos (t+\beta)$, 
respectively, with real constants $t_0$, $\beta$, positive constants $A$, $B$, and 
a real constant $\alpha$ satisfying $|\alpha|<\pi$. Then the density function 
is written as
\be
{\sqrt{2NA\cos \alpha}\over \pi\tilde{\rho}\sqrt{\lambda}}
     \sqrt{1-{A\cos \alpha \over 2N\lambda \tilde{\rho}^2}(x-B\cos(t+\beta))^2}, 
\ee
with 
\be
\tilde{\rho}=\sqrt{\cos^2(t+t_0)+A^2\sin^2(t+\alpha+t_0)}.
\ee
Due to the time-translational invariance, in this case, one of the parameters is simply
related to the time shifting of the density functions.

\subsection{Three-body interaction model}
As another example, we consider the three-body interaction model described by 
the Hamiltonian \cite{GKP}: 
\begin{eqnarray}
H_{3body,s} &=& \sum_{i=1}^3({p_i^2\over 2}+{x_i^2\over 2})  \cr
&&   + \sum_{i>j=1}^3 {\hbar^2\lambda(\lambda-1) \over (x_i-x_j)^2} 
     + 3\sum_{i>j=1}^3{\hbar^2\alpha(\alpha-1) \over y_{ij}^2},
\end{eqnarray}
where $y_{ij}$ is defined as $x_i+x_j-2x_k$ ($k\neq i$ and $k\neq j$).
The (unnormalized) bosonic ground state is given as \cite{GKP}
\be
\phi_{3body}=\prod_{i>j=1}^3(|x_i-x_j|^\lambda |y_{ij}|^\alpha)
             \prod_{i=1}^3 e^{-x_i^2/2\hbar}
\ee
with energy eigenvalue $3\hbar({1\over 2}+(\lambda+\alpha))$. One can easily find 
that the potential of mutual interaction in $H_{3body,s}$ is written in terms of
the differences between positions of two particles.
By applying the formulas in the previous section, one can find that the exact wave 
function for the system described by the Hamiltonian
\begin{eqnarray}
H_{3body,s}&=&\sum_{i=1}^3({p_i^2\over 2M(t)}+M(t)w^2(t){x_i^2\over 2}) \cr
   && + {1 \over M(t)}\sum_{i>j=1}^3 {\hbar^2\lambda(\lambda-1)\over (x_i-x_j)^2}  \cr
   && + {1 \over M(t)}\sum_{i>j=1}^3{\hbar^2\alpha(\alpha-1) \over y_{ij}^2},
\end{eqnarray}
is given as
\begin{eqnarray}
\psi_{3body}^f &=&({u+iv \over \sqrt{\Omega}})^{-3(\lambda+\alpha)-3/2}
       e^{3i\delta_f/\hbar}  \cr
    &&\times \prod_{i=1}^3\exp[{i\over 2\hbar}M{\dot{\rho} \over \rho}(x_i-u_f)^2
          +{i\over \hbar}M\dot{u}_fx_i]\cr
   &&\times(\prod_{i>j=1}^3 |x_i-x_j|^\lambda |y_{ij}|^\alpha)
      \prod_{i=1}^3  e^{-\Omega (x_i-u_f)^2/2\hbar \rho^2}.
\end{eqnarray}

\subsection{Calogero model in Jacobi coordinates}
The model described by the Hamiltonian
\be
\tilde{H}_{C,s}=\sum_{i=1}^N {p_i^2 \over 2} +{1\over 2N}\sum_{i>j=1}^N (x_i-x_j)^2
    +V_C
\ee
has been considered by Calogero \cite{Calogero}, where
\be
V_C=\sum_{i>j=1}^N {\hbar^2 \lambda (\lambda-1) \over (x_i -x_j)^2 }. 
\ee
If one introduce the "Jacobi coordinates"
\begin{eqnarray}
&&y_i = {1\over \sqrt{i(i+1)}}(\sum_{l=1}^i x_l -ix_{i+1}) ~~(i=1,2,\cdots,N-1),\cr
&&y_N = {1\over \sqrt{N}}\sum_{l=1}^N x_l,
\end{eqnarray}
the Hamiltonian in Eq. (41) is written as
\be
\tilde{H}_{C,s}={p_{y_N}^2 \over 2} + H_{C,s},
\ee
where
\be
H_{C,s}= \sum_{i=1}^{N-1} ({p_{y_i}^2 \over 2}+ {y_i^2 \over 2}) +V_C.
\ee
It is easy to see that $V_C$ does not depend on $y_N$ and the Hamiltonian $H_{C,s}$ 
describes a system of interacting $N-1$ particles. In fact, Calogero analyzed
the Hamiltonian system of $H_{C,s}$, and found the (unnormalized) wave functions
\begin{eqnarray}
&&\phi_n^C(y_1,y_2,\cdots,y_{N-1})\cr
   & &=(\prod_{i>j=1}^N(x_i-x_j)^\lambda)
       \exp(-{1\over 2\hbar}\sum_{i=1}^{N-1}y_i^2)
        L_n^b({1\over \hbar}\sum_{i=1}^{N-1}y_i^2)
\end{eqnarray}
satisfying 
\be
H_{C,s} \phi_n^C=E_n\phi_n^C  ~~~~~(n=0,1,2,\cdots),
\ee
where
\begin{eqnarray}
b&=&{1\over 2}(N-3)+{1\over 2}\lambda N(N-1),\\
E_n&=&\hbar[{1\over 2}(N-1)+{1\over 2}\lambda N(N-1)+2n],
\end{eqnarray}
and 
$L_n^b$ is the associated Laguerre polynomials defined through the equation
\be
x{d^2 L_n^b\over dx^2}+(b+1-x){dL_n^b\over dx}+nL_n^b(x)=0.
\ee
If $y_i$ is the space coordinate of the $i$-th particle, the Hamiltonian $H_{C,s}$ does 
{\em not} describe the system of  identical particles, as has been explicitly shown 
in the 3-body system \cite{Cal3body}.

Since the $V_C$ can not be written in terms of $y_i-y_j$, only squeeze-type unitary 
transformation can be applied to give coherent sates. For the system described by the 
Hamiltonian
\be
H_C= \sum_{i=1}^{N-1} ({p_{y_i} \over 2M(t)}+ M(t)w^2(t){y_i^2 \over 2}) 
     +{V_C\over M(t)},
\ee
by making use of the unitary relation in Eq. (14), one may find that
the wave functions satisfying $i\hbar(\partial \psi_n^C/\partial t)=H_C \psi_n^C$
are given as
\begin{eqnarray}
\psi_n^C &=& ({u+iv \over \sqrt{\Omega}})^{-b-1}
         ({u-iv \over \rho})^{2n} \prod_{i>j=1}^N(x_i-x_j)^\lambda     \cr
     &&\times  
     \exp[-{1\over 2\hbar}({\Omega\over \rho^2}-iM{\dot{\rho}\over \rho})
                 \sum_{i=1}^{N-1}y_i^2]
        L_n^b({\Omega\over \hbar\rho^2}\sum_{i=1}^{N-1}y_i^2).
\end{eqnarray}

\section{A Generalization to include external force}
If $V$ can be written in terms of the differences between positions of two particles, 
by modifying the $U_f$, one can find the unitary relation in different Hamiltonian systems, 
as in one-body harmonic oscillator \cite{Song}. By defining the $U_F$ as
\be
U_F= e^{{i \over \hbar} N\delta_F}\prod_{i=1}^N (\exp[{i\over \hbar}M\dot{x}_px_i]
     \exp[-{i\over \hbar}x_p p_i])
\ee 
where $x_p$ and $\delta_F$ are defined through the relations
\begin{eqnarray}
{d \over dt} (M \dot{x}_p) &+& M(t) w^2(t) x_p =F(t),\\
\dot{\delta}_F &=& {1 \over 2}M(w^2 x_p^2 -\dot{x}_p^2),
\end{eqnarray}
one can find the relation
\be
U_F O_N U_F^\dagger= -i\hbar{\partial\over \partial t} + H_{N,F}
\ee
where
\be
H_{N,F}= H_N-F\sum_{i=1}^N x_i.
\ee
From Eq. (56), one can find that the wave function
\begin{eqnarray}
\psi^F
  &=&({\Omega \over \rho^2})^{N/4} 
     ( {u-iv \over \rho} )^{E/\hbar}e^{{i \over \hbar} N\delta_F}  \cr
  &&\times
     ( \prod_{i=1}^N  \exp[{i \over 2 \hbar}M {\dot{\rho} \over \rho} (x_i-x_p)^2 
           +{i\over \hbar}M\dot{x}_px_i] )\cr
  &&\times \phi_s({\sqrt{\Omega} \over \rho}(x_1-x_p),\cdots
           {\sqrt\Omega \over \rho}(x_N-x_p))
\end{eqnarray}
satisfies the Schr\"{o}dinger equation
\be
i\hbar{\partial \psi^F \over \partial t}=H_{N,F}\psi^F.
\ee

\section{Summary and discussions}
It has been shown that the unitary relations in one-body harmonic oscillator 
systems can be extended to give the unitary relations in some of the
Calogero-Sutherland models. These unitary relations can be used not only in finding 
coherent states from the given stationary states in a system, but also in finding 
exact wave functions of the Hamiltonian systems of time-dependent parameters from 
those of time-independent Hamiltonian systems. If the potential of mutual interactions 
can be written in terms of the differences between positions of two particles, 
we have also shown that the wave functions of the system with external force can be 
found from those of the the same system without the external force. 
The list of applications given in this paper is {\em not} exhaustive.

The unitary relations can be formally extended to the case of identical $N$-body free
particles interacting through the mutual interaction potential $V$; however, in this 
case, the $\rho(t)$ diverges as $t^2$ goes to infinity. Even for the case of identical 
$N$-body harmonic oscillators (interacting through the $V$), $\rho(t)$ could be 
unbounded in general, as analyzed in detail for the cases of periodic mass and 
frequencies \cite{SongPRL,SCoherent}.

The system of identical $N$-body free particles interacting through a potential 
${\hbar\lambda(\lambda-1)\pi^2 \over L^2} \sum_{i>j}^N
{1\over \sin^2[\pi(x_i-x_j)/L\sqrt{\hbar}]}$ with a constant $L$ \cite{Sutherland2}, 
has recently been of great interest \cite{SLA,HS,TSA}, while the potential is not of 
homogeneous degree -2. Since the potential satisfies Eq. (2), for the unit mass case, 
if one choose $u_f$ as $at+b$ with constants $a,b$, one can show that the 
Schr\"{o}dinger equation is invariant under a unitary transformation, as in Eq. (22) 
(with $w=0$). For this case, the (unnormalized) wave function of the ground state is given 
as $\psi_0=\prod_{i>j} (\sin{\pi(x_i-x_j) \over L\sqrt{\hbar}})^\lambda$. By applying the
unitary transformation of $U_f$ in Eq. (21) to the $\psi_0$, one can obtain the wave
function $\psi_a=[\prod_{i=1}^N \exp(iax_i /\hbar)]\psi_0$, up to a purely time-dependent
phase. $\psi_a$ is an eigenstate of the Hamiltonian. 
The $\psi_a$ has been discussed as an excited state which may be obtained 
by implementing a Galilei boost to $\psi_0$ \cite{LV}, while the derivation in this paper 
clearly supports this interpretation.

\acknowledgments
This work was supported
in part by grant No. 2000-1-11200-001-2 from the Basic Research Program of the 
Korea Science \& Engineering Foundation.

\end{multicols}

\end{document}